\documentclass[journal]{IEEEtran}
\usepackage{amsmath,amsfonts}
\usepackage{algorithmic}
\usepackage{algorithm}
\usepackage{array}
\usepackage[caption=false,font=normalsize,labelfont=sf,textfont=sf]{subfig}
\usepackage{textcomp}
\usepackage{stfloats}
\usepackage{url}
\usepackage{verbatim}
\usepackage{graphicx}
\usepackage{cite}
\usepackage{booktabs}
\usepackage{color}
\usepackage{multirow}
\usepackage{hyperref}
\begin{document}

\title{Cross-Scope Spatial-Spectral Information Aggregation for Hyperspectral Image Super-Resolution}

\author{Shi Chen, Lefei Zhang, \IEEEmembership{Senior Member, IEEE}, Liangpei Zhang, \IEEEmembership{Fellow, IEEE}
        % <-this % stops a space
\thanks{Manuscript received XXX, revised XXX. This work was supported by the National Natural Science Foundation of China under Grant 62122060 and the Special Fund of Hubei Luojia Laboratory under Grant 220100014. (Corresponding author: Lefei Zhang)}
\thanks{Shi Chen and Lefei Zhang are with the Institute of Artificial Intelligence and School of Computer Science, Wuhan University, Wuhan, 430072, P. R. China. Lefei Zhang is also with the Hubei Luojia Laboratory, Wuhan 430072, P. R. China. ({chenshi, zhanglefei}@whu.edu.cn).}% <-this % stops a space
\thanks{Liangpei Zhang is with the State Key Laboratory of Information Engineering in Surveying, Mapping and Remote Sensing, Wuhan University, Wuhan 430072, P. R. China. (zlp62@whu.edu.cn)}}

\markboth{Journal of \LaTeX\ Class Files,~Vol.~, No.~, FEB~2023}%
{Shell \MakeLowercase{\textit{et al.}}: A Sample Article Using IEEEtran.cls for IEEE Journals}


\maketitle

\begin{abstract}
 Hyperspectral image super-resolution has attained widespread prominence to enhance the spatial resolution of hyperspectral images. However, convolution-based methods have encountered challenges in harnessing the global spatial-spectral information. The prevailing transformer-based methods have not adequately captured the long-range dependencies in both spectral and spatial dimensions. To alleviate this issue, we propose a novel cross-scope spatial-spectral Transformer (CST) to efficiently investigate long-range spatial and spectral similarities for single hyperspectral image super-resolution. Specifically, we devise cross-attention mechanisms in spatial and spectral dimensions to comprehensively model the long-range spatial-spectral characteristics. By integrating global information into the rectangle-window self-attention, we first design a cross-scope spatial self-attention to facilitate long-range spatial interactions. Then, by leveraging appropriately characteristic spatial-spectral features, we construct a cross-scope spectral self-attention to effectively capture the intrinsic correlations among global spectral bands. Finally, we elaborate a concise feed-forward neural network to enhance the feature representation capacity in the Transformer structure. Extensive experiments over three hyperspectral datasets demonstrate that the proposed CST is superior to other state-of-the-art methods both quantitatively and visually. The code is available at \url{https://github.com/Tomchenshi/CST.git}.

%While most window-based Transformers consider regional long-range dependencies in the spatial dimension, they do not take full advantage of global spatial information and spectral correlations.

\end{abstract}


\section{Introduction}
\IEEEPARstart{H}{yperspectral} images provide abundant spectral information on a variety of contiguous spectral bands from the target scene \cite{GS1998}. Unlike traditional imaging techniques that capture color information in limited channels, they possess the remarkable diagnostic ability to distinguish subtle spectral differences and inherent characteristics of various materials \cite{FJ2023}. Therefore, hyperspectral images have been widely employed in diverse fiels, such as target detection \cite{XZ2022, GS2023}, geological exploration \cite{TL2020}, and medical diagnosis \cite{LF2014}. However, an inherent trade-off between spatial and spectral resolution exists in the obtained hyperspectral images due to hardware limitations. Consequently, they frequently suffer from low spatial resolution. Researchers have endeavored to solve these challenges without relying solely on hardware advances.

Super-resolution (SR) seeks to reconstruct high-resolution (HR) images from their low-resolution (LR) counterparts \cite{WC2021}. Recently, the development of SR techniques has extended to hyperspectral images. According to whether auxiliary images are used, hyperspectral image SR can be categorized into fusion-based hyperspectral image super-resolution \cite{DL2019, LH2022, DQ2023} and single hyperspectral image super-resolution (SHSR) \cite{3dfcnn, HL2022}. The fusion-based methods merge the LR hyperspectral images and HR auxiliary images such as multispectral and panchromatic image to reconstruct HR hyperspectral images. Currently, these methods often show superior performance compared to SHSR and have become the dominant approach. Nevertheless, these HR auxiliary images are often difficult to obtain and require precise registration with the target hyperspectral image\cite{ZG2022}. These challenges hinder the implementation of fusion-based methods in practical scenarios.

Without additional auxiliary information, SHSR has a very favourable perspective and it has been widely investigated recently \cite{SSPSR, WL2023}. In the past, traditional methods typically employ handcrafted priors \cite{DF2017, FZ2018, XW2019, GH2021} such as sparse representation, low-rank matrix approximation, and dictionary learning to reconstruct high-resolution hyperspectral images. Nevertheless, these techniques are intricate and inflexible, resulting in limited performance. With the rapid development of deep learning, convolutional neural networks have shown surprising performance in the field of computer vision. By considering the deep SR networks for natural images \cite{EDSR, RDN, RCAN}, hyperspectral image SR has achieved remarkable developments \cite{LL2021}. However, these RGB methods are not directly applicable to hyperspectral images. Some researchers treat hyperspectral images as combinations of different spectra and then employ deep SR networks to process them in a band-by-band manner, which ignores the inherent high-dimensional spectral properties in hyperspectral images. Another intuitive approach is to feed the entire hyperspectral image directly into the deep model for three-channel RGB images, which exponentially increases the parameters and computational cost of the network \cite{3dfcnn}. 

Despite the convolutional neural networks (CNN) tailored for hyperspectral images that have shown good performance \cite{3dfcnn, MCNet}, most of them utilize 2D or 3D convolutions to reveal latent spatial-spectral features in hyperspectral images. Specifically, 2D convolution can only exploit the spatial information in a confined area and is unable to delve into spectral features, which may result the spectral distortion. While 3D convolution can exploit local-range spectral correlations, it cannot model long-range spectral dependencies. Additionally, 3D convolution substantially escalates the network complexity \cite{WL2021}. The grouping-based methods \cite{GDRRN, SSPSR, RFSR} contemplate spectral correlations within local groups. Although the researchers have endeavored to fortify the connections between neighboring spectral groups by designing specialized modules \cite{GELIN}, they have not fully exploited the long-range spectral correlations among global spectra. 

\begin{figure}[t]
\centering
\includegraphics[width=0.488\textwidth]{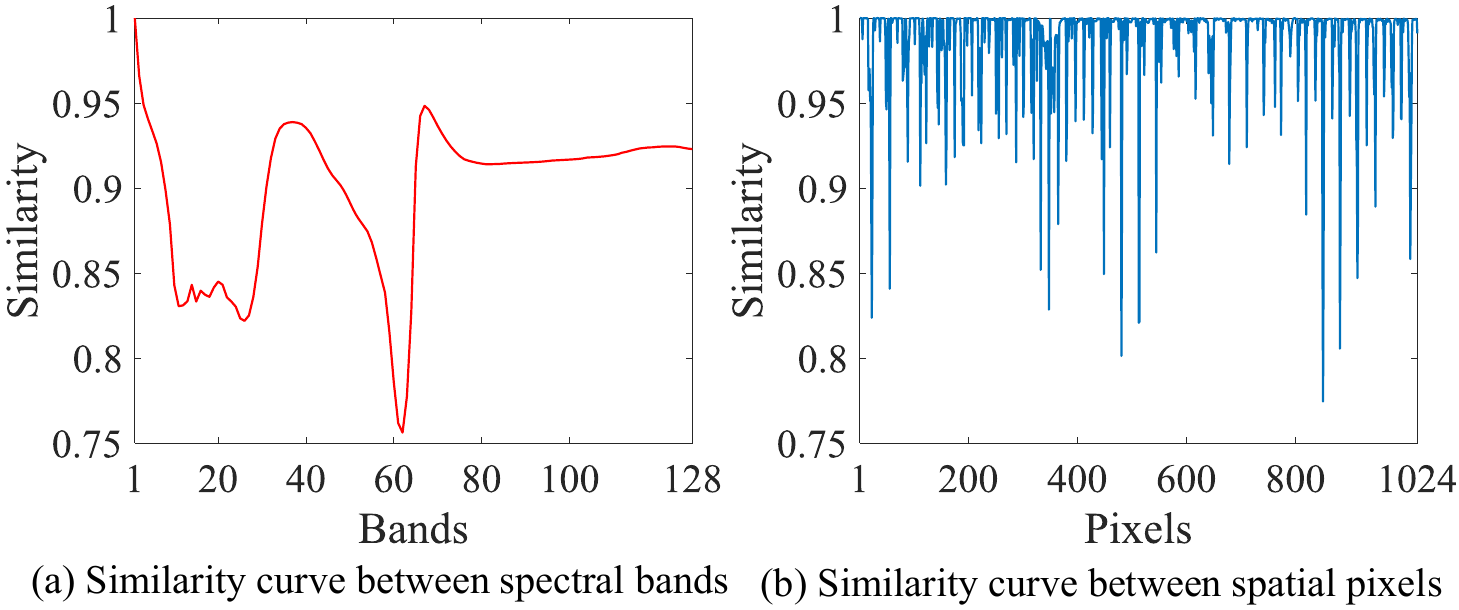}
\caption{The similarity curve in spatial and spectral dimensions. We provide the cosine similarity between the $1$-st spectral band and other spectral bands, as well as between the $1$-st pixel and other pixels in the randomly selected test image.}
\label{fig1}
\end{figure}

Recently, Transformers have become increasingly prevalent in SR tasks due to their capacity to capture long-range dependencies \cite{HH2022, LL2023}. Nevertheless, few researchers have employed dedicated transformers to model global spatial-spectral dependencies for hyperspectral image SR. Evidently, the quadratically growing computational cost with the number of tokens limits the application of Transformer-based methods. Several works for natural images \cite{SWINIR, RESTORMER}, have investigated specific strategies to mitigate this phenomenon, most of which are not compatible with hyperspectral images. The window-based Transformers \cite{SWINIR} limit the long-range spatial information in a regional range, which does not adequately focus on the global spatial dependencies. The channel-wise self-attention \cite{RESTORMER} emphasizes the interactions between high-dimensional channels. Some schemes \cite{HL2022, Interactformer} for hyperspectral image SR adopt the self-attention mechanism in the spectral dimension and leverage 3D convolutions to extract spatial information. However, on the one hand, these methods have not taken into account the long-range spatial dependencies. On the other hand, the low-rank based approaches \cite{ZS2021, GH2021} illustrate that excessive redundancy across high-dimensional channels will adversely impact hyperspectral image SR performance. As shown in Fig. \ref{fig1}, we can first observe in Fig. \ref{fig1} (a) that the first spectral band exhibits strong similarity not only with the adjacent spectral bands but also with the distant spectral bands. Secondly, in Fig. \ref{fig1} (b), the first pixel shows strong similarity with both the neighboring pixels and the pixels at remote positions. Finally, we can notice that in both pixel and spectral curves, there are some regions with relatively weak similarity. Therefore, it can be inferred that there is redundancy in both spatial and spectral dimensions when extracting features. How to effectively model the spatial and spectral long-range dependencies for single hyperspectral image SR remains a significant challenge. 

Motivated by these factors, we propose a cross-scope spatial-spectral Transformer (CST) network to model the long-range spatial and spectral dependencies. In the Transformer module, we individually devise the cross-scope spatial self-attention (CSA) and cross-scope spectral self-attention (CSE). In the spatial dimension, to facilitate the long-range spatial dependencies over the window-based self-attention \cite{SWINTransformer, CAT}, we explore the similarities between local features within rectangular windows and aggregated global features. CSA not only preserves the linear complexity of the window-based attention mechanism but also thoroughly establishes the global dependencies. In the spectral dimension, to alleviate the negative influence of the redundant high-dimensional spectra, we explore the interactions between refined spatial-spectral features and global features. CSE can effectively mitigate the burdens involved in computing long-range spectral dependencies among high-dimensional spectra and achieve the dominant global spectral interactions. Inspired by \cite{RESTORMER}, we construct a concise feed-forward neural network behind the cross-scope self-attentions to reinforce the representation capability of long-range spatial-spectral features. Finally, we aggregate the inductive bias of convolutions with the Transformer layer by merging a parallel channel attention with the cross-scope attentions. Extensive experiments on three common benchmark datasets show that the proposed CST outperforms the state-of-the-art approaches both quantitatively and visually. In summary, the contributions of this paper are summarized as follows:

\begin{itemize}
    \item[1)] We propose a novel cross-scope spatial-spectral Transformer for SHSR to adequately capture the long-range dependencies in both spatial and spectral dimensions, which enhances the SR performance in linear computational complexity.
    \item[2)] To actualize the long-range spatial dependencies, we design a cross-scope spatial self-attention to investigate the similarities between local features within rectangular windows and the aggregated global features.
    \item[3)] To mitigate the effect of redundant high-dimensional spectra and capture the long-range spectral dependencies, we formulate a cross-scope spectral self-attention through the interactions between the characteristic spatial-spectral features and global features.
\end{itemize}

The remaining sections of this paper are organized as follows. Section \uppercase\expandafter{\romannumeral2} briefly reviews existing hyperspectral images SR methods. Section \uppercase\expandafter{\romannumeral3} presents the the proposed method in detail. Section \uppercase\expandafter{\romannumeral4} shows the settings and experimental results and ablation analysis. Finally, Section \uppercase\expandafter{\romannumeral5} concludes this paper.

\section{Related Works}

\subsection{CNN-based single hyperspectral image SR}
 Single hyperspectral image super-resolution has attracted considerable attention owing to its diverse application prospects. Due to the excellent performance of CNNs for natural image SR \cite{EDSR, RDN, RCAN, SAN}, a large number of CNN-based architectures have emerged for addressing hyperspectral image super-resolution. Yuan $et\ al.$ \cite{YZ2017} integrated the transfer learning and collaborative non-negative matrix factorization to guide the reconstruction of hyperspectral image from the natural images. To utilize the high-dimensional spectral characteristics, Mei $et\ al.$ \cite{3dfcnn} proposed a 3D fully CNN network (3DFCNN) to directly extract the spatial-spectral features by the 3D convolutions. To reduce the high computational and memory complexity of 3D convolutions, Li $et\ al.$ \cite{LC2019} decomposed the 3D kernel into the combination of 1D-2D convolutional kernels. Then, Li $et\ al.$ \cite{MCNet} designed a mixed convolutional network (MCNet) to extract the spatial and spectral information by mixing the 2D and separable 3D convolution. Furthermore, Li $et\ al.$ \cite{ERCSR} alternately employed 2D and 3D units to boost the spatial-spectral representation capacity, which fully explored the relationship between 2D/3D convolutions (ERCSR). Then, replacing 3D Convolution with a grouping mechanism, Li $et\ al.$ \cite{GDRRN} further designed a grouped deep recursive residual network (GDRRN) by introducing the group-wise convolution into the recursive residual module. Similarly, Jiang $et\ al.$ \cite{SSPSR} proposed a progressive multi-branch network to learn a spatial-spectral prior of the grouping spectra (SSPSR). To maintain the spatial–spectral consistency, Wang $et\ al.$ \cite{RFSR} proposed a recurrent feedback network (RFSR) with the regularization strategy to strengthen the interaction between the adjacent spectral grouping. Then, Wang $et\ al.$ \cite{GELIN} explored the similar and complementary information within adjacent spectral groups to reconstruct more details. Li $et\ al.$ \cite{dualsr} designed a dual-stage network from the coarse stage to the fine stage to exploit the spatial-spectral similarity between adjacent bands. Nevertheless, these convolution-based methods were restricted to excavating the local spatial-spectral features, which overlooked the long-range spatial-spectral dependencies.

\begin{figure*}[!t]
\centering
\includegraphics[width=0.82\textwidth]{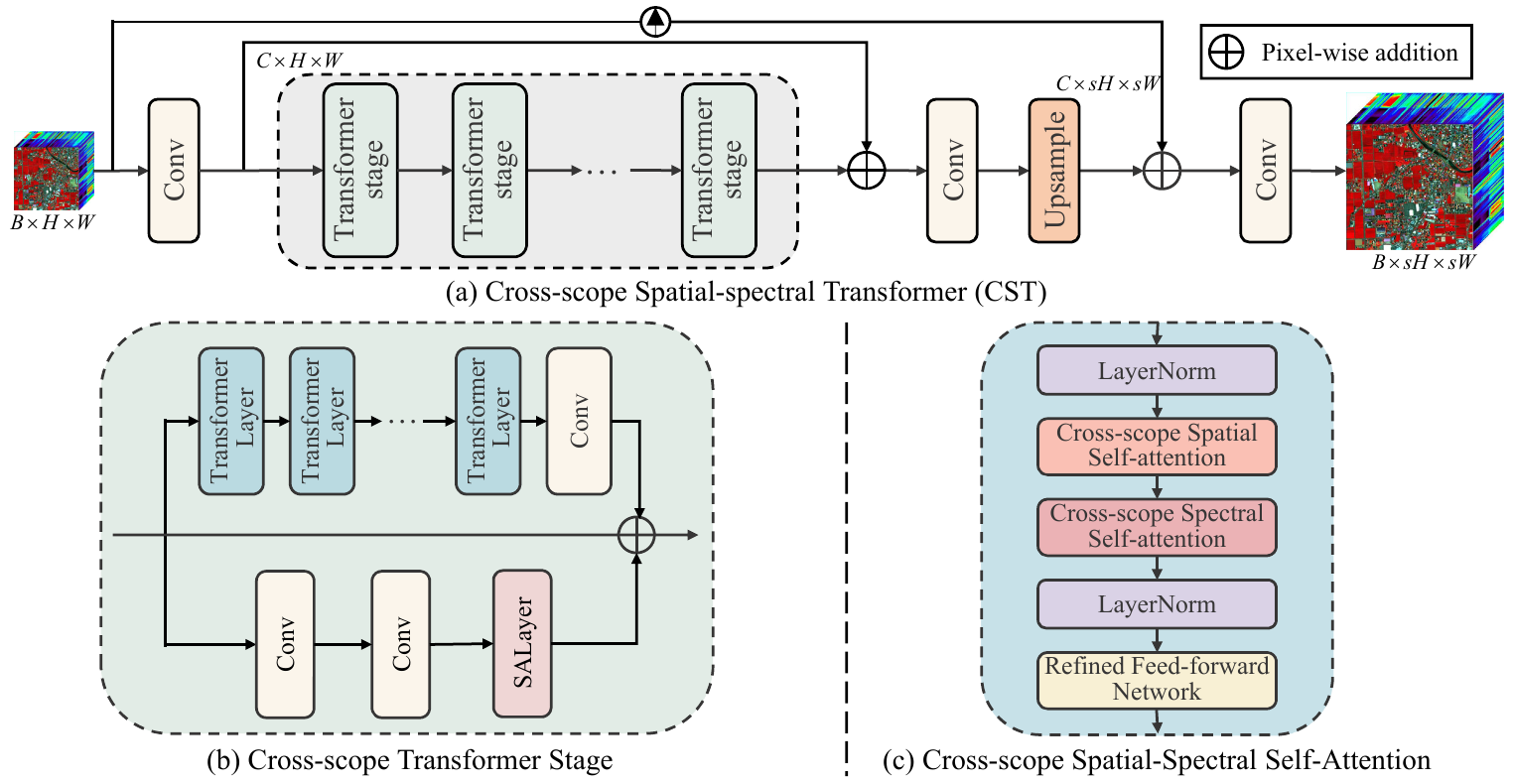}
\caption{The overall architecture of the proposed CST. (a) The main pipeline of CST. (b) The basic cross-scope Transformer block. (c) The cross-scope spatial-spectral self-attention.}
\label{fig2}
\end{figure*}
\subsection{Transformer-based single hyperspectral image SR}
Transformer has received widespread application in SR tasks \cite{CSNL, IPT, RESTORMER} due to its exceptional performance in exploring non-local similarity. Liu $et\ al.$ \cite{Interactformer} designed a parallel branch network called Interactformer, which incorporated both Transformer modules and 3D convolutions. Hu $et\ al.$ \cite{HL2022} proposed a multilevel progressive network (MPNet) to learn the fine details by the progressive learning and non-local channel attention. Wang $et\ al.$ \cite{3DTHSR} also combined the spectrum-wise self-attention and 3D convolutions named 3D-THSR to learn the spatial-spectral features in global receptive field. However, there were several limitations with the aforementioned approach. Firstly, in order to reduce the computational complexity, these methods only explored long-range dependencies in the spectral dimension, thereby failing to fully harness the global spatial information. Secondly, based on the previous analysis, these methods ignored the redundancy when calculating self-attention from the deep features with expanded spectral dimension. Lastly, the utilization of 3D convolutions for extracting local information required substantial memory consumption. Hence, to efficiently capture long-range dependencies in both spectral and spatial dimensions with low complexity, we introduce a cross-scope method to leverage local-global spatial-spectral information.

\section{Proposed Method}
In this section, we first elaborate the overall architecture of the proposed network CST. Subsequently, we present the spatial and spectral cross-scope self-attention mechanisms, and the concise feed-forward neural network, respectively. Finally, we describe the used loss functions in this paper.
\subsection{Overall Architecture}
Motivated by the limitations of current CNNs and Transformer methods in capturing long-range spatial-spectral dependencies \cite{SWINIR, Interactformer}, this paper aims to address the shortcomings of the existing CNN and Transformer in revealing latent spatial-spectral features. While 2D convolutions are confined to local areas and neglect spectral features, 3D convolutions lack the capability to model long-range dependencies and escalate network complexity. Despite the rising popularity of Transformers in SR, few studies have focused on dedicated Transformers for global spatial-spectral dependencies in hyperspectral image SR. Current strategies for natural images are not readily adaptable to hyperspectral data \cite{SSPSR}. The proposed approach seeks to overcome these challenges by introducing a novel cross-scope strategy that leverages local-global spatial-spectral information efficiently.

The overall architecture of the proposed CST is shown in Fig. \ref{fig2}, consisting of three parts: shallow feature extraction, deep feature extraction, and image reconstruction. Given the input LR hyperspectral image $I_{\rm LR} \in \mathbb R^{H \times W \times B}$ and the SR scale factor $s$, where $H$ and $W$ are the height and width in the spatial dimension and $B$ is the number of spectral bands, our network outputs the HR hyperspectral image $I_{\rm SR} \in \mathbb R^{sH \times sW \times B}$. The CST first extracts the shallow features from $I_{\rm LR}$ through a $3\times3$ convolution, which expands the spectral dimensions to obtain more feature maps. The shallow feature extraction can be denoted as
\begin{equation}
    F_{0} = f_{\rm s}(I_{\rm LR}),
\end{equation}
where $f_{\rm s}(\cdot)$ is the shallow feature extraction layer, and $F_{0}$ is the extracted shallow feature. Then, the shallow features pass through a sequence of Transformer stages to extract the deep features. As shown in Fig. \ref{fig2} (b), the designed cross-scope Transformer stage contains successive Transformer layers and one $3\times3$ convolution layer. And each Transformer stage includes two parallel branches, Transformer layer and spectral attention module, to adaptively aggregate the features from the the non-local or local regions. The deep feature extraction can be denoted as 
\begin{equation}
F_{n} = f_{\rm n}(F_{n-1}), n = 1,2,...,N, 
\end{equation}
where $f_{\rm n}(\cdot)$ is the function of $n$-th Transformer stage and $F_{n}$ denotes the $n$-th spatial-spectral feature from the Transformer stage. Subsequently, the deep feature is obtained by the skip connection and one convolution layer, which can be denoted as
\begin{equation}
    F_{d} = Conv(F_{n} + F_{0}),
\end{equation}
where $F_{d}$ represents the extracted deep feature. Finally, the deep feature is fed into the reconstruction layer to improve the spatial resolution, which can be denoted as
\begin{equation}
    F_{up} = F_{\rm up}(F_{d}),
\end{equation}
where $F_{\rm up}(\cdot)$ represents an upsampling operation implemented using pixelshuffle and $F_{up}$ denotes the upsampled feature. The final reconstruction SR image can be attained by long skip connection and one convolution layer, which can be denoted as
\begin{equation}
    I_{SR} = Conv(F_{up} + F_{\rm bi}(I_{LR})),
\end{equation}
where $F_{bi}(\cdot)$ is the bicubic upsampling operation. Through the final residual structure, the network can directly learn the significant textures and details for HR hyperspectral image.

As shown in Fig. \ref{fig2} (c), the designed Transformer layer adopts the common the architecture that consists of some essential blocks: LayerNorm (LN) modules, distinctive cross-scope multi-head self-attention modules, and concise feed-forward neural network. Let $x_l \in \mathbb R^{H \times W \times C}$ denote the input feature of the $l$-th Transformer layer. The outputs of Transformer layer can be expressed as
\begin{equation}
\begin{aligned}
x_{l}' & = CMSA(LN(x_{l-1})) + x_{l-1}\\
x_l & = FFN(LN(x_{l}')) + x_{l}'
\end{aligned}
\end{equation}
where $CMSA(\cdot)$ represents the cross-scope multi-head self-attention, $LN(\cdot)$ represents the function of layer normalization, and $FFN(\cdot)$ represents the designed concise feed-forward neural network. The details of CMSA are shown in Fig.1 (c), which contains a cross-Scope spatial attention and cross-scope spectral attention. 

\subsection{Cross-Scope Spatial Self-attention}
\begin{figure}[h]
\centering
\includegraphics[width=0.48\textwidth]{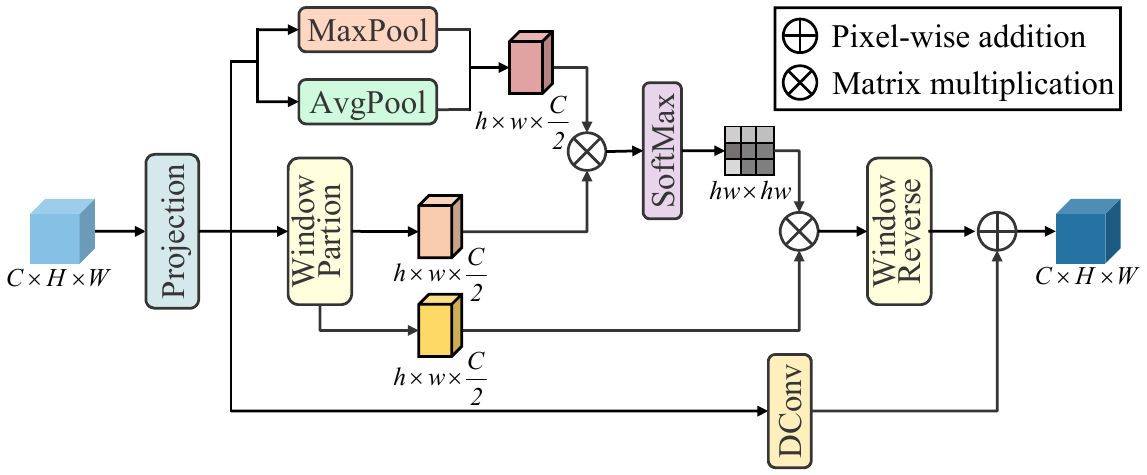}
\caption{The structure of cross-scope spatial self-attention.}
\label{fig3}
\end{figure}
Traditional convolutional approaches struggle to capture extended spatial dependencies, while standard Transformers are constrained by computational complexity. Drawing from these factors, we further introduce the cross-scope spatial self-attention to address the inherent drawback of window-based attention mechanisms that can only account for information within local regions. As shown in Fig. \ref{fig3}, the spatial cross-attention integrates the local and global information within the window-based attention mechanism, aiming to establish comprehensive global spatial dependencies while maintaining linear computational complexity. 

Following \cite{CAT}, for the input feature $X \in \mathbb R^{H \times W \times C}$ after layer normalization, CSA is first divide it into two segments along the spectral dimension, where $X_1 \in \mathbb R^{H \times W \times \frac{C}{2}}$ and $X_2 \in \mathbb R^{H \times W \times \frac{C}{2}}$. Then, $X_1$ and $X_2$ are fed into horizontal and vertical cross-scope spatial multi-head self-attention mechanisms, respectively. Finally, the output $X^{'}$ of CSA is calculated by aggregating the results from the horizontal and vertical dimensions, which can be denoted as
\begin{equation}
\begin{aligned}
X_1, X_2 &= Split(X),\\
X_1^{'} &= W\text{-}CSA(X_1),\ X_2^{'} = H\text{-}CSA(X_2),\\
X^{'} &= CAT(X_1^{'}, X_2^{'}),
\end{aligned}
\end{equation}
where $H\text{-}CSA(\cdot)$ and $W\text{-}CSA(\cdot)$ represents the function of mutli-head self-attention along horizontal and vertical dimensions, respectively. 

Specifically, we calculate the multi-head attention in parallel, and here we take the single head as an example for the sake of simplicity. When the size of rectangle window is $[h,\ w]$ and $h>w$, we partition the feature $X_1$ as $[X_1^{1},X_1^{2},...,X_1^{N}]$ vertically, where $X_1^{i} \in \mathbb R^{H \times W \times \frac{C}{2}}$ and $N=\frac{H\times W}{h \times w}$. Then, the local features are calculated in the vertical windows. To establish long-range spatial dependencies within linear complexity, we directly apply two pooling operations to the input features to obtain global spatial information. Subsequently, the local and global features are reshaped and linearly mapped as query ($Q \in \mathbb R^{hw \times \frac{C}{2}}$), key ($K \in \mathbb R^{hw \times \frac{C}{2}}$), and value ($V \in \mathbb R^{hw \times \frac{C}{2}}$) to compute the cross-attention, which can be denoted as
\begin{equation}
\begin{aligned}
Q_1^{i} = &  X_1^{i}W_Q^{i},\\
K_1^{i} = & CAT(Averagepool(X_1^{i}),Maxpool(x_1^{i})),\\
V_1^{i} = & X_1^{i}W_V^{i},\\
{X_1^{i}}^{'} = & SoftMax(\frac{{Q_1^{i}{{K_1^{i}}^T}}}{\sqrt {\rm{d}}} + P) V),
\end{aligned}
\end{equation}
where $W_Q^{i} K \in \mathbb R^{\frac{C}{2} \times \frac{C}{2}}$ and $W_V^{i} \in \mathbb R^{\frac{C}{2} \times \frac{C}{2}}$ represent the learnable parameters, $Averagepool(\cdot)$ and $Maxpool(\cdot)$ represent the function of adaptive average and max poolings, $B$ is the relative position encoding, and $d$ is the feature dimension. Finally, the output of the vertical CSA is obtained as
\begin{equation}
    X_1^{'} = Merge({X_1^{i}}^{'},{X_2^{i}}^{'},...,{X_1^{N}}^{'}).
\end{equation}

Similarly, for the horizontal cross-scope spatial multi-head self-attention, the size of rectangle window is $[w,h]$ and we can obtain the horizontal output $X_2^{'}$. Additionally, between consecutive cross-scope spatial self-attention, we introduce the shift operation along different directions to enhance interactions between different windows. The previous works have demonstrate that the rectangle windows are advantageous for capturing repetitive textures in various directions. 

The computational complexity of CSA can be denoted as
\begin{equation}
    O(CSA) = (hwHWC),
\end{equation}
where $h$ and $w$ are the constants. In summary, our CSA can calculate the spatial cross-attention with linear computational complexity on the spatial size of $(H\times W)$.

\subsection{Cross-Scope spectral self-attention}
\begin{figure}[h]
\centering
\includegraphics[width=0.48\textwidth]{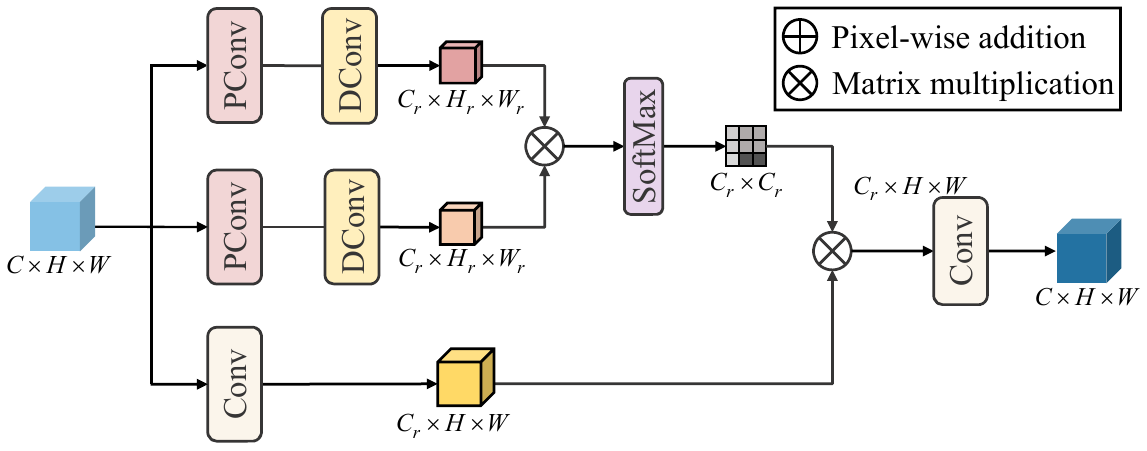}
\caption{The structure of cross-scope spectral self-attention.}
\label{fig4}
\end{figure}
Traditional convolution-based methods adopt the 3D convolution to emphasize local spectral features, disregarding the potential long-range connections. Although some Transformer-based approaches have combined spectral self-attention mechanisms and 3D convolutions, they often suffer from increased memory costs and high complexity due to the high-dimensional feature space. To tackle these issues, we design a cross-scope spectral self-attention mechanism that efficiently captures characteristic spatial-spectral correlations and global interactions. As shown in Fig. \ref{fig4}, the spectral cross-attention models the global spectral dependencies in linear computational complexity.

For the input feature $X^{'} \in \mathbb R^{H \times W\times C}$ from the CSA, CSE first normalizes and reshapes it as $Y \in \mathbb R^{C \times H\times W}$. Then, we sequentially employ the point-wise convolution and depth-wise convolution to extract the characteristic features containing comprehensive spatial-spectral information. We also achieve the multi-head attention mechanism, and here we take the single head as an example for the simplicity. To avoid excessive redundancy across high-dimensional channels, the characteristic features, (i.e. query ($Q \in \mathbb R^{H_r \times W_r \times C_r}$) and key ($K \in \mathbb R^{H_r \times W_r \times C_r}$)), are adaptively calculated through the reduction in both spatial and spectral dimensions. In this paper, we define $H_r = \frac{H}{2}, W_r=\frac{W}{2}$, and $C_r= \frac{C}{2}$. We exploit the cross-attention between the characteristic features and the global feature to achieve the global spatial-spectral interactions, which can be denoted as
\begin{equation}
\begin{aligned}
Q = & W_p^{1}(W_d^{1}(Y)),\\
K = & W_p^{2}(W_d^{2}(Y)),\\
V = & W_v(Y),\\
Y^{'} = & SoftMax(\frac{{Q{{K}^T}}}{\sqrt {\rm{d}}})V),
\end{aligned}
\end{equation}
where $W_p^{(\cdot)}(\cdot)$ is the function of point-wise convolution, $W_d^{(\cdot)}(\cdot)$ is the function of depth-wise convolution, $W_v(\cdot)$ is the standard $1\times1$ convolution, $d$ is the feature dimension, and $Y^{'} \in \mathbb R^{C \times H\times W}$ is the output of CSE. Through CSE, we facilitate the interactions between essential representative features and global features, capturing long-range spectral similarities with low computational cost.

The standard computational complexity of self-attention on spectral dimension is $O(C^{2}HW)$. The computational complexity of the designed CSE can be denoted as
\begin{equation}
    O(CSE)= (C_r^{2}H_{r}W_r),
\end{equation}
where $C_r<C$, $H_r<H$, and $W_r<W$. In summary, our CSA can calculate the spectral cross-attention with low computational cost compared with the original spectral self-attention.

\subsection{Concise Feed-Forward Neural Network}
\begin{figure}[h]
\centering
\includegraphics[width=0.23\textwidth]{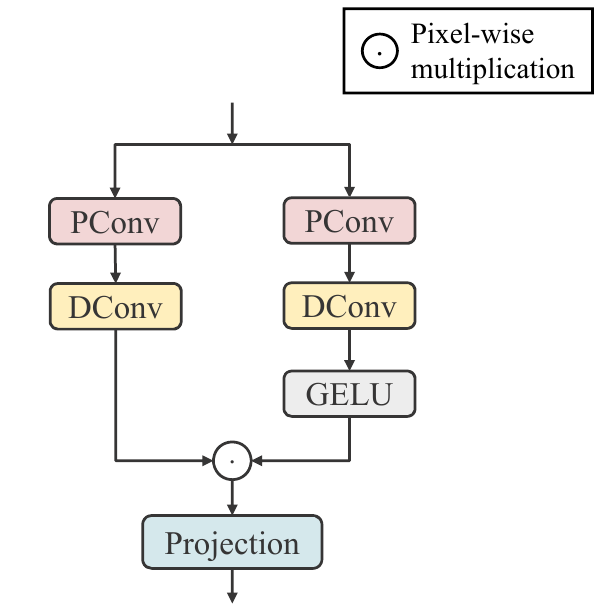}
\caption{The structure of concise feed-forward neural network.}
\label{fig5}
\end{figure}

Traditional feed-forward neural network treat the pixels at each position independently and equally without considering the relationships between them. Following the gating mechanism \cite{RESTORMER}, we design a concise feed-forward neural network (CFN) to improve the representative captivity of the model. As shown in Fig. \ref{fig5}, CFN comprises two parallel branches to process the input spatial-spectral features. In detail, we sequentially utilize point-wise convolution and depth-wise convolution in two branches to extract details beneficial for image reconstruction. Additionally, a non-linear transformation using the GELU activation function is applied in one of the branches, enabling the network to flexibly focus on crucial information. Finally, the advantageous context is learned through the element-wise product of the linear and non-linear branches. Given an input feature $Y^{'} \in \mathbb R^{C \times H\times W}$, the process of CFN can be denoted as
\begin{equation}
    Z^{'} = GELU(W_p^{3}(W_d^{3}(Z))) \odot W_p^{4}(W_d^{4}(Z)).
\end{equation}
Compared with the gating structure in \cite{RESTORMER}, our CFN focuses on exploring internal spatial correlations to uncover more spatial information conducive to image restoration.

\subsection{Loss Function}
Following the previous works, three losses, i.e. $l_{1}$ loss,  spectral angle mapper (SAM) loss, and gradient loss in both spatial and spectral domains, are employed to optimize the proposed network. 

Firstly, the $l_{1}$ loss is widely used in SR for natural image \cite{RCAN}. It calculates the absolute pixel-wise difference between the reconstructed SR image and the original HR image. While the $l_2$ loss tends to generate overly smoothed results, the $l_1$ loss provides a more equitable distribution of errors and enhances the convergence of training. This property encourages the model to reconstruct sharper and more detailed results. In our approach, we adopt the $l_1$ loss to measure the accuracy of the reconstructed hyperspectral images and preserve the spatial details during the super-resolution process. Secondly, The SAM loss is introduced to ensure the spectral consistency of the reconstruction images. Unlike traditional loss functions that focus solely on pixel-wise differences, the SAM loss takes into account the spectral characteristics of hyperspectral data. Finally, inspired by \cite{RFSR, GELIN}, we integrate gradient information to further enhance the sharpness of the reconstructed images. The gradient loss focuses on the differences between adjacent pixels. In conclusion, the total loss for the proposed network can be formulated as 
\begin{align}
&\mathcal{L}_{1}(\theta) = \frac{1}{N}\sum\limits_{n =1}^N {\left\| H_{hr}^n - H_{sr}^n \right\|}_{1},\\
&\mathcal{L}_{sam}(\theta) = \frac{1}{N} \sum\limits_{n =1}^N \frac{1}{\pi} arccos(\frac{H_{hr}^n \cdot H_{sr}^n} {\left\|H_{hr}^n\right\|_{2} \cdot \left\|H_{sr}^n\right\|_{2}}),\\
&\mathcal{L}_{gra}(\theta) = \frac{1}{N} \sum\limits_{n =1}^N \left\| M(H_{hr}^n) - M(H_{sr}^n)) \right\|_{1},\\
&\mathcal{L}_{total}(\theta) = \mathcal{L}_{1} + \lambda_s\mathcal{L}_{sam} + \lambda_g\mathcal{L}_{gra},
\end{align}
where $N$ is the number of the input in a training batch, $H_{hr}^n$ and $H_{sr}^n$ represents the $n$-th HR and SR hyperspectral images, and $M(\cdot)$ represents the horizontal, vertical, and spectral gradients of the input, respectively. $\lambda_s$ and $\lambda_g$ denote the hyper-parameters that control the balance between the different losses for excellent SR performance. The analysis of the loss functions is shown in the section of the experiment. In this paper, $\lambda_s$ is set to 0.3 and $\lambda_g$ is set to 0.1, empirically.

\section{Experiments and Results}

\subsection{Datasets}
The experiments are conducted on three common real-scenario hyperspectral image datasets, i.e., Chikusei \cite{chikusei}, Pavia Center \cite{pavia}, and Houston \footnote{https://hyperspectral.ee.uh.edu/?page\_id=1075}.

\subsubsection{Chikusei Dataset}
The Chikusei dataset is a hyperspectral remote sensing image, acquired by the Headwall Hyperspec-VNIR-C imaging sensor over agricultural and urban areas in Chikusei, Ibaraki, Japan. The image contains 128 spectral bands ranging from 363 to 1018 nm with a spatial resolution of $2517 \times 2335$. The ground sampling distance is 2.5 m.

\subsubsection{Houston Dataset}
The Houston 2018 dataset is the part of the 2018 IEEE
GRSS Data Fusion Contes, acquired by the ITRES CASI 1500
spectral imager over the University of Houston campus and
its surrounding urban area in Houston, Texas, America. The image contains 48 spectral bands ranging from 380 to 1050 nm with a spatial resolution of $4172 \times 1202$. The ground sampling distance is 1 m. 

\subsubsection{Pavia Centre Dataset }
The Pavia Centre dataset is a hyperspectral remote sensing image, taken by the Reflective Optics
System Imaging Spectrometer (ROSIS) sensor during a flight
campaign over the center area of Pavia, northern Italy, in 2001. The image contains 103 bands ranging from 430 to 860 nm with a spatial resolution of $1096 \times 1096$. The ground sampling distance is 1.3 m.

\subsection{Implementation Details}
 In the proposed network CST,  we set the kernel size of the standard convolution and depth-wise convolution as $3 \times3$, and the size of others for point-wise convolution and spectral attention is $1 \times1$. Zero padding is adopted to maintain the spatial size of feature maps. Following the settings in \cite{RCAN}, we set the reduction ratio in CA as 16. We set the number of channels $C$ to $180$, the number of Transformer stages as $4$, and the number of Transformer layers as $4$. In general, the rectangle window size is set to $2\times 16$.  In the image reconstruction, we adopt the progressive upsampling strategy by PixelShuffle \cite{SC2016} to upsample the input LR hyperspectral images for reducing the parameters (e.g., upsampling 3 times for scale factor $ \times$8). For the loss function, we set $\lambda_1 = 0.3$ and $\lambda_2=0.1$. In the training procedure, the Adam optimizer with the default setting is employed to train the network for 500 epochs, and the size of the mini-batch is set to 32. The initial learning rate is set to $1e^{-4}$ and halves every 100 epochs until 300 epochs. The proposed model is implemented by Pytorch on NVIDIA RTX 3090 GPU.

 We compare the proposed CST with other state-of-the-art methods at three scale factors, including one interpolation method, one Transformer-based SR SwinIR \cite{SWINIR} for natural image and deep learning-based SHSR methods, i.e. Bicubic, GDRRN \cite{GDRRN}, SSPSR \cite{SSPSR}, RFSR \cite{RFSR}, and GELIN \cite{GELIN}. For SwinIR, We directly feed the whole hyperspectral image into the network for training and modify the dimensions of the input and output. For the aforementioned methods, we strive to achieve their optimal performance. Six popular evaluation metrics are used for the comparison experiments to entirely demonstrate the performance of the models in both spatial and spectral dimensions, including peak signal-to-noise ratio (PSNR), structure similarity (SSIM), spectral angle mapper (SAM), cross correlation (CC), root-mean-squared error (RMSE), and erreur relative global adimensionnellede synthese (ERGAS). While PSNR, SSIM, and RMSE are typically employed to assess the quality of natural image restoration, the remaining three metrics, CC, SAM, and ERGAS, are common evaluation measures in hyperspectral image fusion tasks.

\subsection{Experiments on the Chikusei dataset}

\begin{figure*}[t]
\centering
\includegraphics[width=0.9\textwidth]{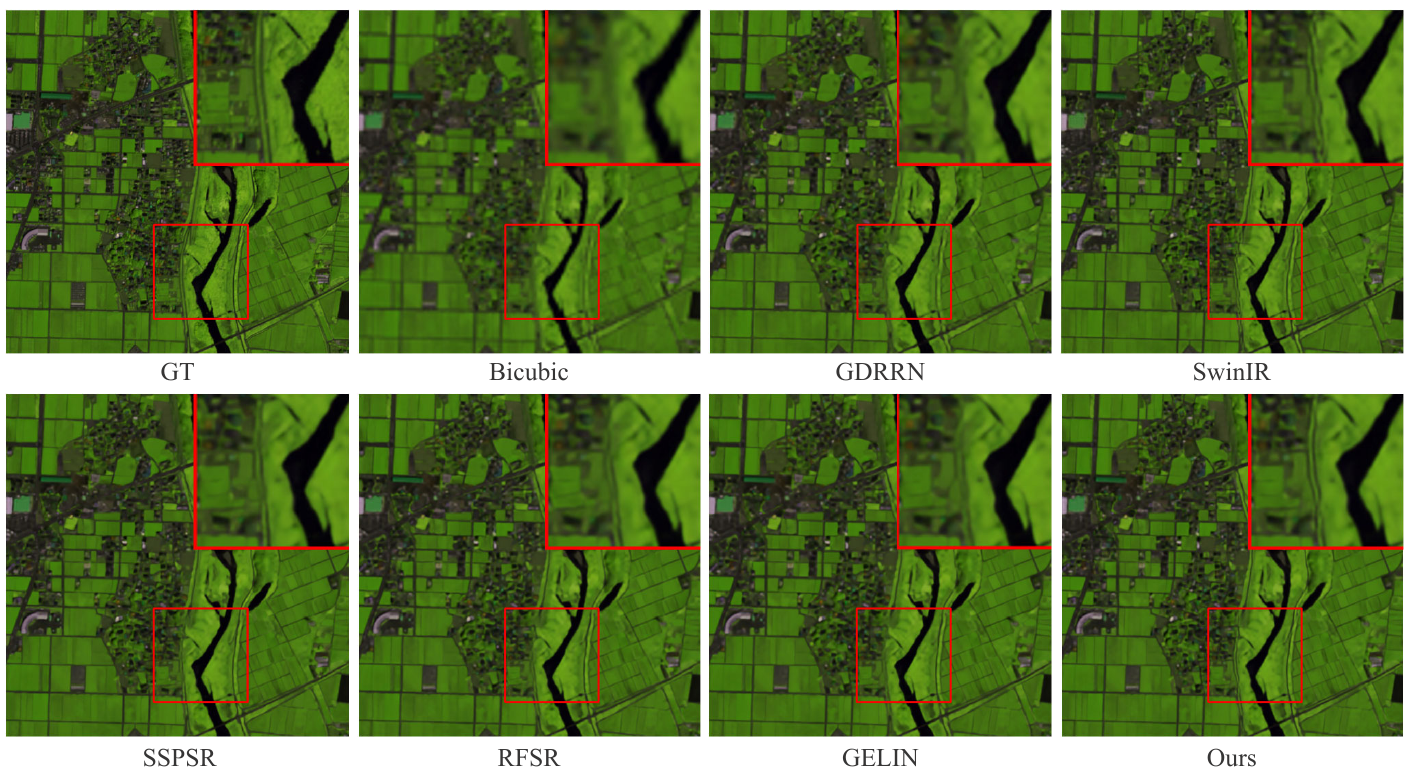}
\caption{Reconstructed test hyperspectral image in the Chikusei dataset with spectral bands 70-100-36 as R-G-B with the scale factor $\times$4. (From Left to Right) Ground truth, results of Bicubic, GDRRN \cite{GDRRN}, SwinIR\cite{SWINIR}, SSPSR \cite{SSPSR}, RFSR \cite{RFSR}, GeLIN\cite{GELIN}, and the proposed CST.}
\label{fig6}
\end{figure*}

\begin{figure*}[!t]
\centering
\includegraphics[width=0.9\textwidth]{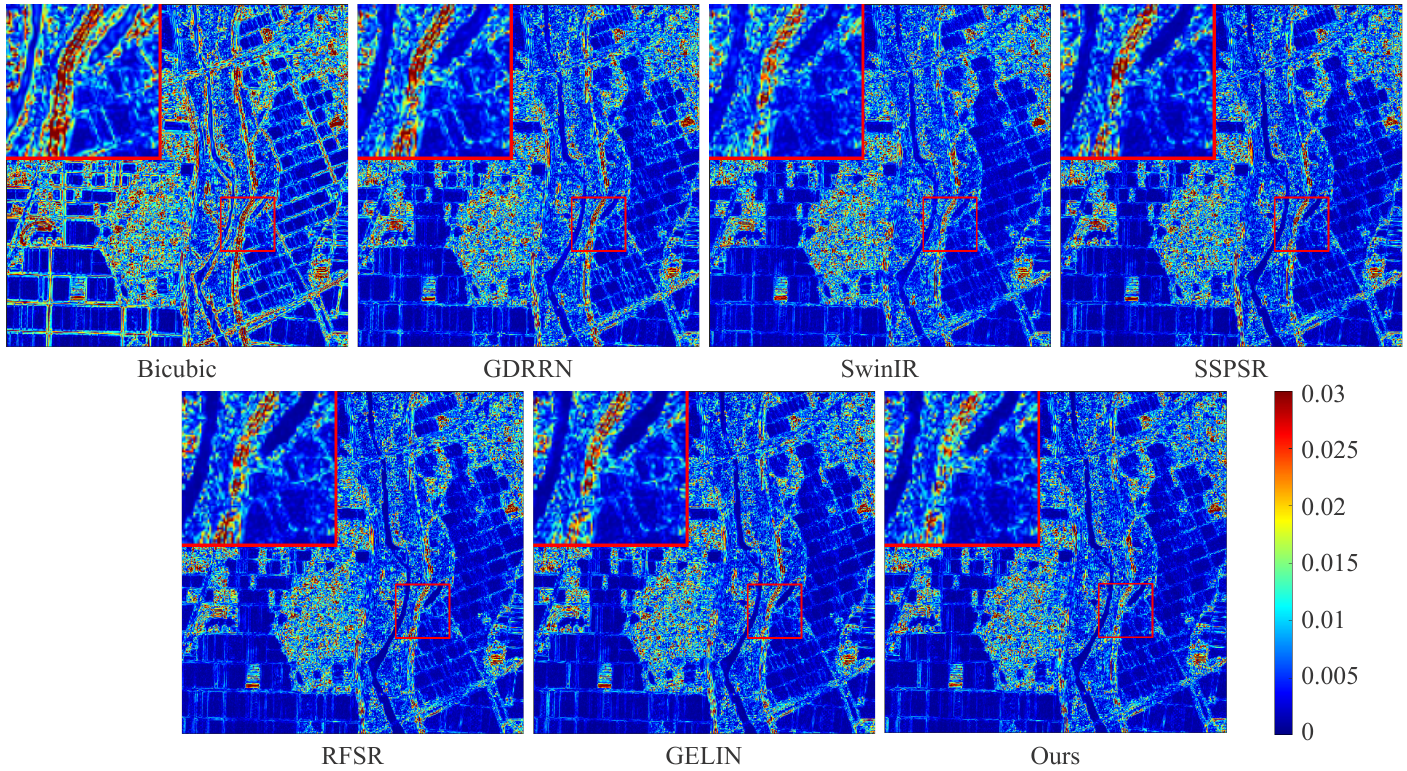}
\caption{Error maps of the test hyperspectral image on the Chikusei dataset at the scale factor $\times$4.}
\label{fig7}
\end{figure*}

\begin{figure*}[!t]
\centering
\includegraphics[width=0.8\textwidth]{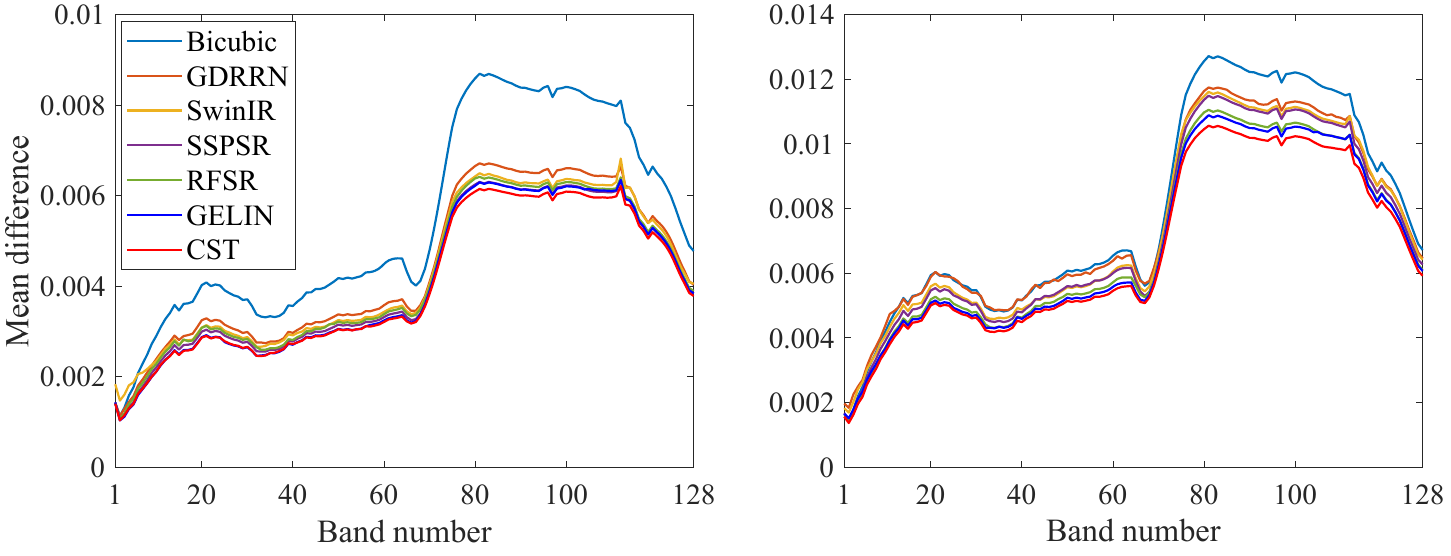}
\caption{Mean spectral difference curve of hyperspectral testing images on the Chikusei dataset at the scale factors $\times$4 (Left) and $\times$8 (Right), respectively.}
\label{fig8}
\end{figure*}

\begin{table}[h]
\caption{Quantitative performance on the Chikusei dataset at different scale factors. Bold represents the best result and underline represents the second best}
\label{tab1}
\centering
\setlength{\tabcolsep}{0.009\hsize}{
\begin{tabular}{cccccccc}
\toprule[1pt]
Method & Scale & PSNR$\uparrow$    & SSIM$\uparrow$   & SAM$\downarrow$    & CC$\uparrow$     & RMSE$\downarrow$   & ERGAS$\downarrow$  \\
\midrule
\midrule
bicubic & $\times2$ & 43.2125 & 0.9721 & 1.7880 & 0.9781 & 0.0082 & 3.5981 \\
GDRRN \cite{GDRRN}  & $\times2$ & 46.4286 & 0.9869 & 1.3911 & 0.9885 & 0.0056 & 2.6049 \\
SwinIR \cite{SWINIR} & $\times2$ & 47.3018 & 0.9889 & 1.2228 & 0.9898 & 0.0051 & 2.4146 \\
SSPSR \cite{SSPSR} & $\times2$ & 47.4073 & 0.9893 & \underline{1.2035} & 0.9906 & 0.0051 & 2.3177 \\
RFSR  \cite{RFSR}  & $\times2$ & \underline{47.6258} & \underline{0.9898} & 1.1335 & \underline{0.9910} & \underline{0.0049} & \underline{2.2862} \\
Gelin \cite{GELIN}  & $\times2$ & -       & -      & -      & -      & -      & -      \\
Ours    & $\times2$ & \textbf{48.0852} & \textbf{0.9908} & \textbf{1.1057} & \textbf{0.9918} & \textbf{0.0047} & \textbf{2.1685} \\

\midrule
\midrule
bicubic & $\times4$ & 37.6377          & 0.8954          & 3.4040          & 0.9212          & 0.0156          & 6.7564          \\
GDRRN \cite{GDRRN}  & $\times4$ & 39.0864          & 0.9265          & 3.0536          & 0.9421          & 0.0130          & 5.7972          \\
SwinIR \cite{SWINIR} & $\times4$ & 39.5366          & 0.9364          & 2.8327          & 0.9456          & 0.0122          & 5.6280          \\
SSPSR \cite{SSPSR}  & $\times4$ & 39.9797          & 0.9393          & 2.4864          & 0.9528          & 0.0119          & 5.1905          \\
RFSR \cite{RFSR}   & $\times4$ & 39.8950          & 0.9382          & 2.4656          & 0.9517          & 0.0120          & 5.2334          \\
Gelin \cite{GELIN}  & $\times4$ & \underline{40.1573}    & \underline{0.9410}    & \underline{2.4266}    & \underline {0.9543}    & \underline {0.0118}    & \underline{5.0314}    \\
Ours    & $\times4$ & \textbf{40.2406} & \textbf{0.9431} & \textbf{2.3453} & \textbf{0.9554} & \textbf{0.0116} & \textbf{5.0123} \\
\midrule
\midrule

bicubic & $\times 8$ & 34.5049 & 0.8069 & 5.0436 & 0.8314 & 0.0224 & 9.6975 \\
GDRRN \cite{GDRRN}  & $\times 8$ & 34.7395 & 0.8199 & 5.0967 & 0.8381 & 0.0213 & 9.6464 \\
SwinIR \cite{SWINIR} & $\times 8$ & 34.8785 & 0.8307 & 5.0413 & 0.8465 & 0.0210 & 9.4743 \\
SSPSR \cite{SSPSR}  & $\times 8$ & 35.1643 & 0.8299 & 4.6911 & 0.8560 & 0.0206 & 9.0504 \\
RFSR \cite{RFSR}   & $\times 8$ & 35.5049 & 0.8405 & 4.2785 & 0.8661 & 0.0199 & 8.6338 \\
Gelin \cite{GELIN}  & $\times 8$ & \underline{35.6496} & \underline{0.8464} & \underline{4.1354} & \underline{0.8707} & \underline{0.0197} & \underline{8.4520} \\
Ours    & $\times 8$ & \textbf{35.7902} & \textbf{0.8522} & \textbf{3.9915} & \textbf{0.8753} & \textbf{0.0192} & \textbf{8.3636} \\
\bottomrule[1pt]
\end{tabular}
}
\end{table}

After removing the blurred edges, the size of the remaining central region in the Chikusei dataset is $2304\times2048\times128$. Following the previous work \cite{SSPSR, GELIN}, the four non-overlapping images with the size of $512\times 512\times 128$ are cropped from the top region for testing. The remaining area is cropped into overlapping HR images for training (10\% of the training data is randomly selected for validation). Specifically, the spatial resolution of LR training patches is $32\times32$, and the sizes of corresponding HR patches for scale factors $\times2$, $\times4$, and $\times8$ are $64\times64$, $128\times128$, and $256\times256$, respectively. All the LR patches from the hyperspectral image are generated through bicubic downsampling at different scales. 

 The quantitative results of our method and other compared methods at different scale factors on the Chikusei dataset are shown in Table \ref{tab1}. The best results are indicated in bold, and the second-best results are underlined. We can observe that the interpolation-based method Bicubic exhibits ordinary SR performance, while learning-based methods achieve significant improvements. While the Transformer-based method SWINIR can slightly enhance spatial recovery results, this approach designed for natural image super-resolution does not consider the spectral characteristics of hyperspectral images, leading to spectral distortion. Grouping-based methods designed for hyperspectral images restore the HR images in both spatial and spectral dimensions but do not explore the non-local spatial-spectral similarities. Notably, through capturing the global spatial-spectral dependencies, our proposed CST achieves the best performance for all metrics at scale factors $\times2$, $\times4$, and $\times8$, which demonstrates the effectiveness and superiority of our method.

 As shown in Fig. \ref{fig6}, the qualitative results of all methods at scale factor $\times 4$ on the Chikusei dataset are presented. Especially, the 70th, 100th, and 36th bands of the test image are selected as the R-G-B channels for visual comparison. We can observe that while interpolation-based method Bicubic upsampling can achieve hyperspectral image SR, it often introduces blur and overly smooth details. Recent methods like SSPSR and GELIN have made progress in restoring overall structure, but they still suffer from unclear boundaries and severe artifacts. In particular, the visual results demonstrate that the proposed CST can reconstruct the HR hyperspectral images with clearer and sharper details compared with the other methods. Moreover, the advantages of our method can be better observed based on the enlarged view of the details in the red box. Additionally, we also visualize the mean error images among all spectra in Fig. \ref{fig7} to provide an intuitive view of the accuracy of the reconstruction for individual pixels. In the error images, a bluer color indicates a higher accuracy for the reconstruction results. According to the highlighted red box region in error maps, our method exhibits minimal errors, which illustrates the superiority of the proposed method. Finally, the mean spectral difference curves of the test images at the scale factors $\times 4$ and $\times 8$ from the testing dataset are shown in Fig. \ref{fig8} to evaluate the SR results from a spectral perspective. Compared to analyzing the reflectance of random pixels in the reconstructed image, the mean spectral difference offers a comprehensive measure of the accuracy of spectral reconstruction quality across the entire image. The lower the curve, the fewer spectral differences and distortion between the SR results and the ground-truth. Clearly, our method achieves the best spectral reconstruction results at different scale factors.

\subsection{Experiments on the Houston dataset}

\begin{figure*}[t]
\centering
\includegraphics[width=0.9\textwidth]{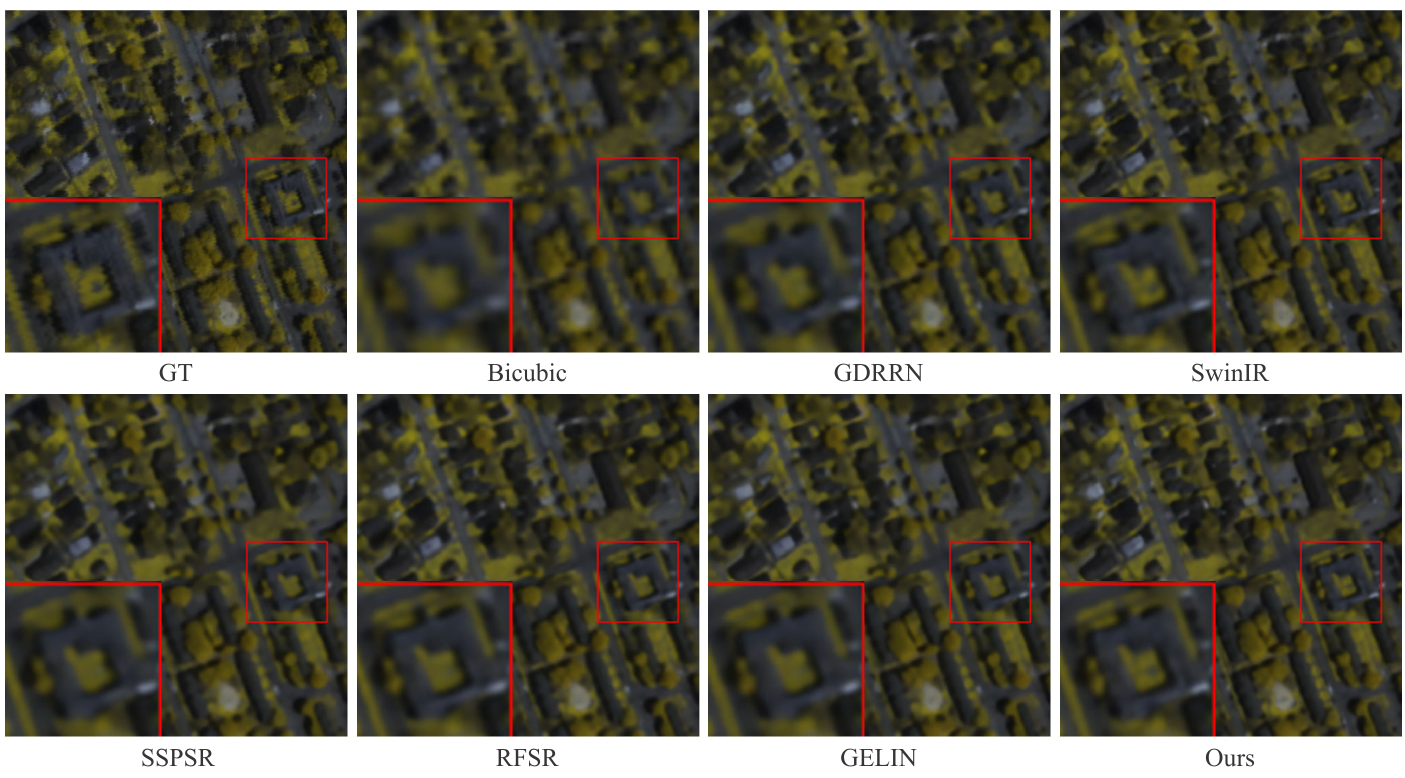}
\caption{Reconstructed test hyperspectral image on the Houston dataset with spectral bands 29-26-19 as R-G-B with the scale factor $\times$4. (From Left to Right) Ground truth, results of Bicubic, GDRRN \cite{GDRRN}, SwinIR\cite{SWINIR}, SSPSR \cite{SSPSR}, RFSR \cite{RFSR}, Gelin\cite{GELIN}, and the proposed CST.}
\label{fig9}
\end{figure*}

% \begin{figure*}[!t]
% \centering
% \includegraphics[width=0.99\textwidth]{fig10.pdf}
% \caption{Error maps of the test hyperspectral image on the Houston dataset at the scale factor $\times$4.}
% \label{fig10}
% \end{figure*}

% \begin{figure*}[!h]
% \centering
% \includegraphics[width=0.8\textwidth]{fig11.pdf}
% \caption{Mean spectral difference curve of hyperspectral testing images on the Houston dataset at the scale factors $\times$4 (Left) and $\times$8 (Right), respectively.}
% \label{fig11}
% \end{figure*}

\begin{table}[h]
\caption{Quantitative performance on the Houston dataset at different scale factors. Bold represents the best result and underline represents the second best}
\label{tab2}
\centering
\setlength{\tabcolsep}{0.009\hsize}{
\begin{tabular}{cccccccc}
\toprule[1pt]
Method & Scale & PSNR$\uparrow$    & SSIM$\uparrow$   & SAM$\downarrow$    & CC$\uparrow$     & RMSE$\downarrow$   & ERGAS$\downarrow$  \\
\midrule
\midrule
bicubic & $\times2$ & 49.4735 & 0.9915 & 1.2707 & 0.9940 & 0.0040 & 1.3755 \\
GDRRN \cite{GDRRN}  & $\times2$ & 51.5205 & 0.9949 & 1.1241 & 0.9957 & 0.0031 & 1.0723 \\
SwinIR \cite{SWINIR} & $\times2$ & 52.2118 & 0.9957 & 1.0863 & 0.9964 & 0.0028 & 0.9898 \\
SSPSR \cite{SSPSR} & $\times2$ & 52.1246 & 0.9954 & 1.0101 & 0.9965 & 0.0029 & 1.0035 \\
RFSR  \cite{RFSR}  & $\times2$ & \underline{52.5131} & \underline{0.9958} & \underline{0.9491} & \underline{0.9968} & \underline{0.0028} & \underline{0.9571} \\
Gelin \cite{GELIN}  & $\times2$ & -       & -      & -      & -      & -      & -      \\
Ours    & $\times2$ & \textbf{53.4784} & \textbf{0.9967} & \textbf{0.8803} & \textbf{0.9974} & \textbf{0.0025} & \textbf{0.8570} \\

\midrule
\midrule
bicubic & $\times4$ & 43.0272          & 0.9613          & 2.5453          & 0.9741          & 0.0086          & 2.9085          \\
GDRRN \cite{GDRRN}  & $\times4$ & 44.2964          & 0.9730          & 2.5347          & 0.9760          & 0.0069          & 2.4700          \\
SwinIR \cite{SWINIR} & $\times4$ & 46.0971          & 0.9808          & 1.9463          & 0.9864          & 0.0059          & 2.0039           \\
SSPSR \cite{SSPSR}  & $\times4$ & 45.6017          & 0.9778          & 1.9650          & 0.9850          & 0.0063          & 2.1380          \\
RFSR \cite{RFSR}   & $\times4$ & 45.8677          & \underline{0.9792}          & \underline{1.8304}          & 0.9858          & \underline{0.0060}         & \underline{2.0659}          \\
Gelin \cite{GELIN}  & $\times4$ & \underline{45.8715}    & 0.9790    & 1.8759    & \underline {0.9859}    & 0.0061    & 2.0778    \\
Ours    & $\times4$ & \textbf{46.9957} & \textbf{0.9839} & \textbf{1.6508} & \textbf{0.9892} & \textbf{0.0054} & \textbf{1.8236} \\
\midrule
\midrule

bicubic & $\times 8$ & 38.1083 & 0.8987 & 4.6704 & 0.9177 & 0.0152 & 5.1229  \\
GDRRN \cite{GDRRN}  & $\times 8$ & 38.2592 & 0.9085 & 4.9045 & 0.9138 & 0.0140 & 4.9135 \\
SwinIR \cite{SWINIR} & $\times 8$ & 39.4013 & 0.9194 & 4.0586 & 0.9370 & 0.0127 & 4.3333 \\
SSPSR \cite{SSPSR}  & $\times 8$ & 39.2844 & 0.9164 & 4.2673 & 0.9346 & 0.0130 & 4.4212 \\
RFSR \cite{RFSR}   & $\times 8$ & 39.4899 & 0.9211 & 3.8403 & 0.9379 & 0.0126 & 4.2967 \\
Gelin \cite{GELIN}  & $\times 8$ & \underline{39.6387} & \underline{0.9216} & \underline{3.9231} & \underline{0.9393} & \underline{0.0125} & \underline{4.2453} \\
Ours    & $\times 8$ & \textbf{39.9786} & \textbf{0.9261} & \textbf{3.5802} & \textbf{0.9442} & \textbf{0.0121} & \textbf{4.0958} \\
\bottomrule[1pt]
\end{tabular}
}
\end{table}

Following the above settings, the eight non-overlapping images from the Houston dataset with the size of $256\times 256\times 48$ are cropped from the top region for testing. The remaining area is cropped into overlapping HR patches for training (10\% of the training data is randomly selected for validation). Specifically, the spatial resolution of LR training patches is $32\times32$, and the sizes of corresponding HR patches for scale factors $\times2$, $\times4$, and $\times8$ are $64\times64$, $128\times128$, and $256\times256$, respectively. 

 In Table \ref{tab2}, the quantitative results of all methods at different scale factors on the Houston dataset are shown. It can be observed that our CST method outperforms other methods for all metrics at different scale factors. Moreover, compared to the results on other datasets, our method shows greater improvements on the Houston dataset. This is attributed to the Transformer-based structure of our method, which requires a substantial amount of training data to enhance the capabilities of the model, and the Houston dataset offers more training samples. The results further substantiate the effectiveness of our method, and emphasize the significance of long-range spatial and spectral similarities in hyperspectral image SR.

Similarly, as shown in Fig. \ref{fig9}, we also present the qualitative results of all methods at scale factor $\times 4$ on the Houston dataset. Specifically, the 29th, 26th, and 19th bands of the test image are treated as the R–G–B channels for visual comparison. It can be observed that other methods inevitably produce many blurry edges and artificial artifacts, while our method is capable of recovering more low-frequency and high-frequency information in the SR hyperspectral images. Moreover, the enlarged view of the details in the red box presents the advantages of our method. 

% In addition, we also visualize the error maps of all methods at factor scale $\times 4$ on the testing image in Fig. \ref{fig10}. Similar to the results on the Chikusei dataset, previous methods like GDRRN \cite{GDRRN} and SSPSR \cite{SSPSR} have more red pixel regions compared to other methods. Although RFSR \cite{RFSR} and GELIN \cite{GELIN} have fewer red pixels than them, our proposed MSDformer provides the least number of red pixels and reconstruction errors among all algorithms (please refer to the area marked with red boxes). Finally, the mean spectral difference curves of the test images at the scale factors $\times 4$ and $\times 8$ from the Houston testing dataset are shown in Fig. \ref{fig11} to evaluate the spectral consistency. Since the spectral pattern in the Houston dataset is relatively simple compared to the other datasets, our method also achieves excellent performance on all metrics, which further verifies the strengths of our approach. 

\subsection{Experiments on the Pavia dataset}

\begin{figure*}[t]
\centering
\includegraphics[width=0.9\textwidth]{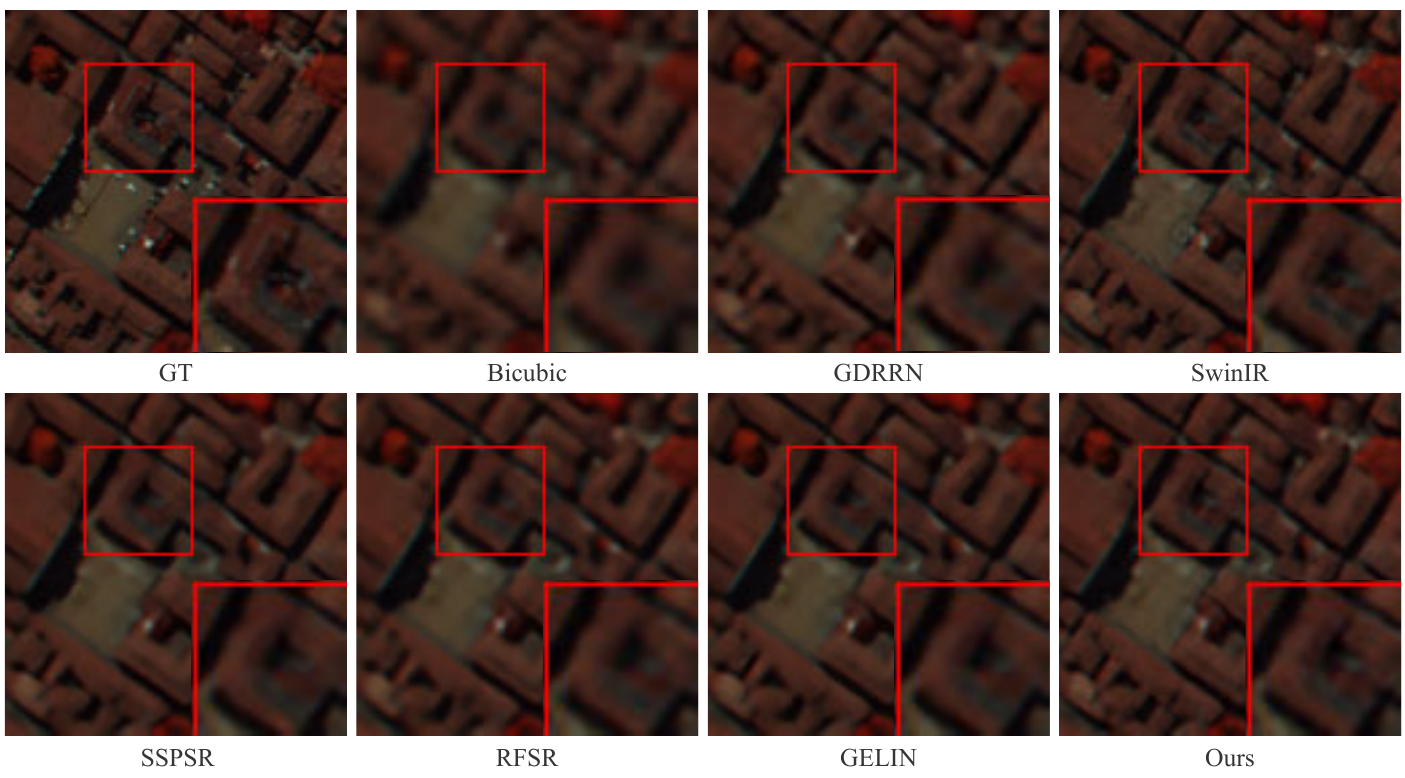}
\caption{Reconstructed test hyperspectral image on the Pavia dataset with spectral bands 100-30-12 as R-G-B with the scale factor $\times$4. (From Left to Right) Ground truth, results of Bicubic, GDRRN \cite{GDRRN}, SwinIR\cite{SWINIR}, SSPSR \cite{SSPSR}, RFSR \cite{RFSR}, Gelin\cite{GELIN}, and the proposed CST.}
\label{fig12}
\end{figure*}

% \begin{figure*}[!t]
% \centering
% \includegraphics[width=0.99\textwidth]{fig13.pdf}
% \caption{Error maps of the test hyperspectral image on the Pavia dataset at the scale factor $\times$4.}
% \label{fig13}
% \end{figure*}

% \begin{figure*}[!h]
% \centering
% \includegraphics[width=0.8\textwidth]{fig14.pdf}
% \caption{Mean spectral difference curve of hyperspectral testing images on the Pavia dataset at the scale factors $\times$4 (Left) and $\times$8 (Right), respectively.}
% \label{fig14}
% \end{figure*}

\begin{table}[t]
\caption{Quantitative performance on the Pavia dataset at different scale factors. Bold represents the best result and underline represents the second best}
\label{tab3}
\centering
\setlength{\tabcolsep}{0.009\hsize}{
\begin{tabular}{cccccccc}
\toprule[1pt]
Method & Scale & PSNR$\uparrow$    & SSIM$\uparrow$   & SAM$\downarrow$    & CC$\uparrow$     & RMSE$\downarrow$   & ERGAS$\downarrow$  \\
\midrule
\midrule
bicubic & $\times2$ & 32.5489          & 0.9143          & 4.4086          & 0.9491          & 0.0242          & 4.0984  \\
GDRRN \cite{GDRRN}  & $\times2$ & 34.3982         & 0.9484          & 4.1618          & 0.9681          & 0.0182          & 3.1716 \\
SwinIR \cite{SWINIR} & $\times2$ & \underline{36.6070}          & 0.9620          & 3.4767          & 0.9765          & \underline{0.0151}          & 2.6903 \\
SSPSR \cite{SSPSR} & $\times2$ & 36.5800          & \underline{0.9627}          & \underline{3.4717}          & \underline{0.9766}          & 0.0152          & \underline{2.6856} \\
RFSR \cite{RFSR}  & $\times2$ & 35.7468 & 0.9564 & 3.6104 & 0.9724 & 0.0167 & 2.9409  \\
Gelin \cite{GELIN}  & $\times2$ & -       & -      & -      & -      & -      & -      \\
Ours    & $\times2$ & \textbf{37.3427} & \textbf{0.9677} & \textbf{3.2840} & \textbf{0.9795} & \textbf{0.0139} & \textbf{2.5000} \\

\midrule
\midrule
Bicubic & $\times4$ & 27.8623 & 0.7240 & 6.1325 & 0.8522 & 0.0424 & 6.8670 \\
GDRRN \cite{GDRRN}   & $\times4$ & 28.7845 & 0.7889 & 6.6320 & 0.8776 & 0.0377 & 6.2168 \\
SwinIR \cite{SWINIR} & $\times4$ & 29.5134 & \underline{0.8227} & 5.6553 & 0.8966 & 0.0346 & 5.7325 \\
SSPSR \cite{SSPSR}  & $\times4$ & \underline{29.5355} & 0.8223 & 5.4462 & \underline{0.8975} & \underline{0.0346}  & \underline{5.7045} \\
RFSR \cite{RFSR}   & $\times4$ & 29.4013 & 0.8152 & 5.5216 & 0.8938 & 0.0351 & 5.8014 \\
Gelin \cite{GELIN}  & $\times4$ & 29.4854          &0.8204          & \underline{5.4265}         & 0.8967          & 0.0348          &5.7317    \\
Ours    & $\times4$ & \textbf{29.7451} & \textbf{0.8311} & \textbf{5.3885} & \textbf{0.9021} & \textbf{0.0337} & \textbf{5.5829} \\
\midrule
\midrule

bicubic & $\times 8$ & 24.9874          & 0.4744          & 7.6196          & 0.6892          & 0.0603          & 9.5021  \\
GDRRN \cite{GDRRN}  & $\times 8$ & 25.1226          & 0.5030          & 8.3504          & 0.6931          & 0.0591          & 9.3617 \\
SwinIR \cite{SWINIR} & $\times 8$ & 25.3301          & 0.5229          & 7.9841          & 0.7102          & 0.0577          & 9.1353 \\
SSPSR \cite{SSPSR}  & $\times 8$ & 25.5094          & 0.5410          & 7.4538          & \underline{0.7242}          & 0.0566          & 8.9497  \\
RFSR \cite{RFSR}   & $\times 8$ & \underline{25.4987}          & 0.5332          & 7.3518          & 0.7227          & \underline{0.0565}          & \underline{8.9615} \\
Gelin \cite{GELIN}  & $\times 8$ & 25.4975          & \underline{0.5418}          & \underline{7.3091}          &0.7234          & 0.0566          & 8.9651 \\
Ours    & $\times 8$ & \textbf{25.5692} & \textbf{0.5450} & \textbf{7.2958} & \textbf{0.7284} & \textbf{0.0561} & \textbf{8.8889} \\
\bottomrule[1pt]
\end{tabular}
}
\end{table}
After removing the noise region without information, the size of the available part in the Pavia dataset is $1096\times 715\times 102$. For scale factors $\times2$ and $\times 4$, the four non-overlapping images with the size of $256\times 256\times 102$ are cropped from the left region for testing. Additionally, for the scale factor $\times 8$, the non-overlapping images with the size of $128\times 128\times 102$ are cropped from the left region for testing. The remaining area is cropped into overlapping HR images for training (10\% of the training data is randomly selected for validation). Specifically, the spatial resolution of LR training patches is $16\times16$, and the sizes of corresponding HR patches for scale factors $\times2$, $\times4$, and $\times8$ are $32\times32$, $64\times64$, and $128\times128$, respectively. 

In Table \ref{tab3}, the quantitative results of all methods at different scale factors on the Houston dataset are shown. Compared to the other datasets, the Pavia dataset has a lower spatial resolution. Consequently, it is a challenge for our Transformer-based method CST due to the limited training samples. In addition, all methods perform poorly relatively on this dataset. Moreover, Nonetheless, our method still manages to outperform other methods across all metrics on this dataset. In summary, our method achieves competitive performance on the dataset with limited samples, which demonstrates the effectiveness of non-local spatial and spectral similarities. 

As shown in Fig. \ref{fig12}, we  visualize the qualitative results of all methods from the testing image at the scale factor ×4. Especially, the 100th, 30th, and 12th bands of the testing image are selected as the R–G–B channels for visual comparison. The observations for the previous dataset are also applicable to the Pavia dataset. Similarly, other methods generate unclear contours and textures, while our approach can restore more fine details. Based on the region in the red box, we can observe the superiority of our approach. 

% Finally, we also visualize the error maps and mean spectral difference curves of the test image in Fig. \ref{fig13} and Fig. \ref{fig14}. According to the analysis, all of these results illustrate the effectiveness of our proposed method.

\subsection{Ablation study}
In this section, we provide the performance of different variants at the scale factor $\times 4$ on Chikusei dataset to verify the effectiveness of our method. We calculate the model parameters and the floating point operations (Flops) of all methods when the size of input is $32\times32\times128$.

\subsubsection{Effectiveness of the proposed cross-scope self-attention}
\begin{table*}[t]
\caption{Abaltion Study. Quantitative comparisons of some variants of the proposed method \\over the Chikusei testing dataset at scale factor $\times$4}
\label{tab4}
\centering
\begin{tabular}{ccccccccc}
\toprule[1pt]
Variant              & Params.($\times 10^{6}$) & FLOPs($\times 10^{9}$)    & PSNR$\uparrow$    & SSIM$\uparrow$   & SAM$\downarrow$    & CC$\uparrow$     & RMSE$\downarrow$   & ERGAS$\downarrow$  \\
\midrule
w/o CSA     & 8.992& 19.135& 40.1703 & 0.9420 & 2.3736 & 0.9547 & 0.0116 & 5.0560   \\
w/o CSE     & 10.062& 20.223& 39.9529 & 0.9391 & 2.4104 & 0.9524 & 0.0119 & 5.1921   \\
CSA-CSA     & 12.189& 22.376& 40.0703 & 0.9408 & 2.3742 & 0.9536 & 0.0118 & 5.1140   \\
CSE-CSE     & 10.049& 20.199& 40.0298 & 0.9407 & 2.4209 & 0.9533 & 0.0118 & 5.1556   \\
CSA-CSE(Ours) & 11.119& 21.287& \textbf{40.2406} & \textbf{0.9431} & \textbf{2.3453} & \textbf{0.9554} & \textbf{0.0116} & \textbf{5.0123}   \\
\bottomrule[1pt]
\end{tabular}
\end{table*}

In the proposed cross-scope multi-head self-attention, we devise a cross-scope spatial self-attention and a cross-scope spectral self-attention to model the spatial and spectral long-range dependencies separately. To verify the effectiveness of the proposed CST, CSA, and the structure of the cross-scope self-attention, we design various variants of CST. By removing CSA or CSE independently, we denote the variants as ``Ours w/o CSA" or ``Ours w/o CSE". To investigate the impact of self-attention mechanisms in different dimensions, we employ two consecutive CSAs by replacing CSE with CSA in the variant and denote it as ``Ours with 2CSA". Similarly, another variant containing two consecutive CSEs is denoted as ``Ours with 2CSE". As shown in Table \ref{tab4}, when using a single CSA or CSE, the performance of the model significantly deteriorates. When using two consecutive CSAs, the PSNR decreases by dB compared to ``w/o CSA". The possible reason is the excessive redundancy in exploring spatial information, making it difficult for the model to learn effectively. Compared with the other variants, our model achieves the best performance, which demonstrates the higher efficiency
of the proposed cross-scope multi-head self-attention.

\begin{table*}[t]
\caption{Abaltion Study. Quantitative comparisons about the reduction strategy in CSA \\over the Chikusei testing dataset at scale factor $\times$4}
\label{tab5}
\centering
\begin{tabular}{ccccccccc}
\toprule[1pt]
Variant              & Params.($\times 10^{6}$) & FLOPs($\times 10^{9}$)    & PSNR$\uparrow$    & SSIM$\uparrow$   & SAM$\downarrow$    & CC$\uparrow$     & RMSE$\downarrow$   & ERGAS$\downarrow$  \\
\midrule
w/o spe     & 12.170& 22.352& 40.2147 & 0.9427 & 2.3633 & 0.9551 & 0.0116 & 5.0321   \\
w/o spa     & 11.133& 21.311& 40.1936 & 0.9425 & 2.3627 & 0.9550 & 0.0116 & 5.0481   \\
Ours        & 11.119& 21.287 &\textbf{40.2406} & \textbf{0.9431} & \textbf{2.3453} & \textbf{0.9554} & \textbf{0.0116} & \textbf{5.0123}  \\
\bottomrule[1pt]
\end{tabular}
\end{table*}

\begin{table*}[ht]
\caption{Abaltion Study. Quantitative comparisons of different sizes of the rectangle window in CSA over Chikusei testing dataset at scale factor $\times$4. Bold represents represents the best}
\label{tab6}
\centering
\begin{tabular}{ccccccccc}
\toprule[1pt]
Window Size & Params.($\times 10^{6}$) & FLOPs($\times 10^{9}$) & PSNR$\uparrow$    & SSIM$\uparrow$   & SAM$\downarrow$ & CC$\uparrow$     & RMSE$\downarrow$   & ERGAS$\downarrow$\\
\midrule
(2,2)  &11.119  &21.286  & 40.1888 & 0.9425 & 2.3719 &0.9549 &0.0116 & 5.0403\\
(2,4)  &11.119  &21.286  & 40.1374 & 0.9421 & 2.3883 &0.9544 &0.0117 &5.0881\\
(2,8)  &11.119  &21.287  & 40.1756 & 0.9424 & 2.3596 &0.9548 &0.0116 &5.0558\\
(2,16) & 11.119& 21.287& \textbf{40.2406} & \textbf{0.9431} & \textbf{2.3453} & \textbf{0.9554} & \textbf{0.0116} & \textbf{5.0123}\\
\bottomrule[1pt]
\end{tabular}
\end{table*}

\subsubsection{Effectiveness of the reduction strategy}
As we analyze in Fig. \ref{fig1} in the introduction, there is a significant amount of redundancy in both spatial and spectral dimensions, which can be a computational burden when calculating self-attention. Therefore, in CSA, we reduce the computation of features in both spatial and spectral dimensions. Additionally, we study the effect of this dimensionality reduction strategy in Table \ref{tab5}. We can see that when not using dimensionality reduction in either spatial or spectral dimensions, the network's computation and complexity increase significantly, and its performance does not improve but, in fact, slightly decreases. This demonstrates the effectiveness of our reduction strategy.

\subsubsection{Analysis of the rectangle window in CSA}

In order to investigate the impact of different sizes of rectangular windows in CSA, we conducted a large amount of experiments with various sizes of rectangular windows. Table \ref{tab6} presents the relationship between the window size and the experimental results. We can observe that as the spatial self-attention window size increases, our proposed network can capture more spatial information in larger regions, and both spatial and spectral metrics improve accordingly without extra computational cost. This demonstrates the importance of long-range spatial dependencies in hyperspectral image SR. Therefore, for better results, we adopt a rectangle window with the size of (2, 16) in this study.

\subsubsection{Effectiveness of different losses}
\begin{table}[!t]
\caption{Quantitative performance (average PSNR/SSIM/SAM) of different loss functions evaluated over four testing \\images of Chikusei dataset at the scale factor $ \times 4$. Bold represents represents the best}
\label{tab7}
\centering
\begin{tabular}{cccccc}
\toprule[1pt]
$\mathcal{L}_{1}$ & $\mathcal{L}_{spe}$ & $\mathcal{L}_{gra}$ & PSNR$\uparrow$ & SSIM$\uparrow$   & SAM$\downarrow$  \\ 
\midrule
\checkmark  &            &              & 40.0777 & 0.9413 & 2.4751 \\
\checkmark  & \checkmark &              & 40.1939 & 0.9425 & 2.3579 \\
\checkmark  &            & \checkmark   & 40.1229 & 0.9418 & 2.4951 \\
\checkmark  & \checkmark & \checkmark   & \textbf{40.2406} & \textbf{0.9431} & \textbf{2.3453} \\ 
\bottomrule[1pt]
\end{tabular}
\end{table}

We study the effectiveness of different losses in our experiments. The performance of different combinations of all losses is shown in Table \ref{tab7}. When considering only the $l_{1}$ loss, our method achieves the poorest results, especially in terms of PSNR and SSIM. By only incorporating the single spectral loss, our approach obtains the second-best results in both spatial and spectral metrics, indicating that leveraging spectral information can enhance performance and assist in the construction of HR hyperspectral image in spatial and spectral dimensions. Moreover, with only the gradient loss, our method improves the PSNR in terms of spatial dimensions while sacrificing the spectral accuracy. We can conclude that the gradient loss enhances the sharpness of high-resolution images, resulting in clearer details. When all loss functions are employed, our method achieves the best results, suggesting the effectiveness of using multiple loss functions in our approach. Under the constraints of the loss functions used in this paper, our method effectively explores global spectral similarity.

\subsubsection{Analysis of the number of Transformer stages}
\begin{table}[!t]
\caption{Quantitative comparisons of the number of Cross-scope Transformer stages over the Chikusei testing dataset at scale factor $\times$4. Bold represents represents the best}
\label{tab8}
\centering
\begin{tabular}{cccc}
\toprule[1pt]
Number ($N$) & PSNR$\uparrow$    & SSIM$\uparrow$   & SAM$\downarrow$ \\
\midrule
$N=2$ & 40.1189 & 0.9416 & 2.3967  \\
$N=4$ & \textbf{40.2406} & \textbf{0.9431} & \textbf{2.3453}  \\
$N=6$ & 40.2312 & 0.9430 & 2.3523  \\
$N=8$ & 40.2080 & 0.9428 & 2.3404  \\
\bottomrule[1pt]
\end{tabular}
\end{table}

Our CST model consists of multiple consecutive Transformer stages, and here we investigate the impact of the number of Transformer stages $N$ on the experimental results in Table \ref{tab8}. It can be observed that when using fewer modules ($N = 2$), our method produces the worst results. As we increase $N$ to 4 and 6, the quantitative metrics improve accordingly. However, when we further set $N$ to 8, the experimental results deteriorate. The main reason is that as the network depth increases, the Transformer architecture requires more data for training, and the network is prone to overfitting with poor generalization ability.

\section{Conclusion}
In this study, we propose a new method CST to addresses the challenges in hyperspectral image super-resolution. In essence, CST primarily leverages the designed cross-scope spatial-spectral self-attention to effectively capture both long-range spatial and spectral dependencies. Through aggregating the global representative features with the rectangle window self-attention mechanism in the spatial dimension, the devised CSA efficiently explores the interactions between the local and global features in linear complexity. By introducing the reduction strategy in spectral self-attention, the designed CSE significantly integrates the comprehensive features and refined spatial-spectral features, which reduces the computational cost without sacrificing the SR performance. Moreover, we adopt a concise feed-forward neural network to further process the long-range spatial-spectral features, which can enhance the representational capacity and generalization of the network. Finally, we demonstrate the effectiveness and superiority of CST through extensive experiments on various datasets. CST consistently outperforms state-of-the-art methods quantitatively and visually, thanks to its ability to capture long-range spatial and spectral similarities.

% \section*{Acknowledgments}
% This should be a simple paragraph before the References to thank those individuals and institutions who have supported your work on this article.

\bibliographystyle{IEEEtran}
\bibliography{ref}

% \newpage

 

% \end{IEEEbiography}



% \vfill

\end{document}